\documentclass[prl]{revtex4}

\usepackage{graphicx}
\usepackage{dcolumn}
\usepackage{bm}

\begin{document}
\renewcommand{\theequation}{\arabic{equation}}

\title{Perspectives on gravity-induced radiative processes in astrophysics\\}

\author{Giorgio Papini}
\altaffiliation[Electronic address:]{papini@uregina.ca}
\affiliation{Department of Physics and Prairie Particle Physics
Institute, University of Regina, Regina, Sask, S4S 0A2, Canada}


\begin{abstract}

Single-vertex Feynman diagrams represent the dominant
contribution to physical processes, but are frequently forbidden kinematically.
This is changed when the particles involved propagate in a gravitational
background and acquire  an effective mass. Procedures are introduced
that allow the calculation of lowest order diagrams,
their corresponding transition probabilities, emission powers and spectra to all
orders in the metric deviation, for
particles of any spin propagating in gravitational fields described by any metric.
Physical properties of the "space-time medium" are also discussed. It is shown in particular
that a small dissipation term in the particle wave equations
can trigger a strong back-reaction that introduces resonances in the radiative process
and affects the resulting gravitational background.

\end{abstract}

 \maketitle

\setcounter{equation}{0}
\section{Introduction}

The search for particle processes of astrophysical significance to
which gravitation makes a non-negligible contribution is in
general difficult, though potentially rewarding. The entire field
of gravitational lensing is in fact based on the discovery of one
such process in which light interacts with gravity \cite{ehlers}. It is hoped that a
number of additional processes will point out new directions of investigation.
The search is made more difficult,
unfortunately, by the fact that the lowest order Feynman diagrams
that represent the dominant contribution to a
process, are frequently forbidden by kinematics. Consider, for
instance, the process in which an incoming massive particle of
momentum $p_{\mu}$ and dispersion relation $p_{\mu}p^{\mu}
=m_{1}^2$ produces a photon of momentum $\ell_{\mu}\ell^{\mu}=0$,
while the outgoing particle has momentum $p'_{\mu}p'^{\mu}=m^2$.
Conservation of energy-momentum requires
$p_{\mu}=p'_{\mu}+\ell_{\mu}$. In the rest frame of $p$ we have
$0=\vec{p'}+\vec{\ell}$, which gives $\vec{\ell}=-\vec{p'}$,
$p_{0}=m_{1}$ and again $m_{1}=\sqrt{\ell^2+m^2}+\ell$. Then
$(m_{1}-\ell)^2=\ell^2 +m^2$ leads to $\ell=(m_{1}^2-m^2)/2m_{1}$
which shows that for $m_{1}=m$, the case considered, we get $\ell=0$ and the process
becomes physically meaningless. There are processes, however, in
which massive particles emitting a photon are not kinematically
forbidden. This is certainly the case when
gravitation alters the dispersion relations of at least one of the
particles involved \cite{PRD82}. We stress here that even the reduction of a Feynman
diagram by a single vertex would result in a cross-section
gain of a factor $(GM/R)^2$, where $M$ and $R$ are mass and radius of
the gravitational source (units $\hbar=c=1$).

External gravitational fields have long been known to play the role of a
medium \cite{Land,Peters} in the propagation of particles, be these treated classically, or
according to quantum mechanics. In the latter case scattering by a Newtonian potential
has been the subject of several investigations \cite{bocc, papvall, overhauser}, but
bremsstrahlung \cite{dewitt, dehnen, audretsch, bezerra}, the emission of
C\^erenkov radiation \cite{gupta} and other processes \cite{ahluw}
have also been studied in connection with various gravitational sources.
As stated above, external gravitational fields do alter the dispersion relations
of a particle propagating in a gravitational background at least
to the extent that the particle is no longer on shell. It would therefore appear sufficient
to somehow solve a wave equation to obtain the desired result. This is done
in quantum electrodynamics where the electromagnetic field
is however represented by a four-vector of which only a single component is usually
taken into account \cite{jauch}.
The case of gravity, represented by a second rank tensor, is considerably more complicated.
Moreover, the theoretical prediction by Mashhoon \cite{mashhoon1, mashhoon2, mashhoon3, mashhoon4},
confirmed by other authors \cite{hehl.ni, caipap, papprd1,papprd2},
of the existence of a coupling of gravity to spin, requires that the effect of a gravitational field
be no longer limited to a single component of the metric. This has become even more pressing since the experimental
observation of spin-rotation coupling for photons \cite{ashby} and neutrons \cite{demirel} and of other
important spin-induced effects at the macroscopic level \cite{Everitt, Iorio1, Iorio2}.

External gravitational fields contribute to the solution of
covariant wave equations through a Berry phase \cite{caipap0,GRG22}. This should be
expected because in metric theories of gravitation \cite{MPLA}, general
relativity in particular, the space parameter of Berry's theory
coincides with space-time. It has been shown that the wave equations for fermions and
bosons can be solved exactly to first order in the metric
deviation $\gamma_{\mu\nu}=g_{\mu\nu}-\eta_{\mu\nu}$ for any
metric $g_{\mu\nu}$ and that the phases so calculated \cite{caipap,papa,papb,dinesh,pap3} give reliable
results in
interferometry, gyroscopy \cite{caipap0} and optics \cite{pap2,pap4}, give the correct
Einstein deflection, can be used in the study of neutrino helicity
and flavour oscillations \cite{pap1} and of spin-gravity coupling in general \cite{papa,pap3}.
They also reproduce a variety of known effects as discussed
in \cite{papb,dinesh} and \cite{zitter}.

The dispersion relations of a particle propagating in a
gravitational background  can be derived from the respective
covariant wave equations. The gravitational phases mentioned above,
change, in effect, a particle four-momentum by acting on the wave
function of the field-free equations. This result applies equally
well to fermions and bosons and can be extended to all orders in
$\gamma_{\mu\nu}$ \cite{MPLA}. The calculation of even the most elementary
Feynman diagrams does require an appropriate treatment
when gravitational fields are present.
The procedures developed
in \cite{PRD82} fill in part this gap and apply to linearized
gravitational, or inertial fields of any type up to intermediate
intensities. The present paper focuses on the applications
of the procedures outlined in \cite{PRD82} rather than on the individual reactions,
though the aim still is to find physical processes capable of leading to potentially observable results.
Below, we give examples of processes that can be treated
in ways that are similar. Because of the intrinsic resemblance, gravitational bremsstrahlung \cite{Peters1,Carmeli,Galtsov,Kovacs,Misner}
will be added to this category in due time.

Additional interesting
possibilities do moreover arise when particles propagate in a gravitational background.
In the conclusions, for instance, we briefly discuss properties of
the "space-time medium" such as dispersion.

\section{The process $ P \rightarrow p'+\gamma $}

\begin{figure}
\centering
\includegraphics[width=0.3\textwidth]{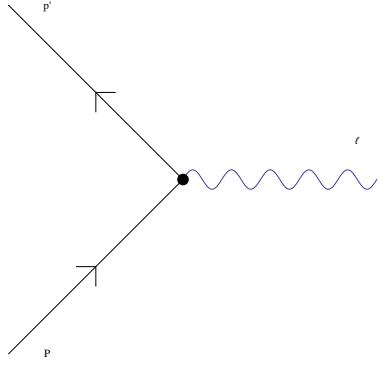}
\caption{\label{fig:Feynman2} $p'$ and $\ell$ are the outgoing fermion and
photon and $ P$ indicates the incoming fermion.}
\end{figure}

Let us assume, for simplicity, that $P$ in Fig.1 is an incoming
fermion and that the photon $\ell$ and outgoing fermion $p'$ are
produced on-shell. The solution of the covariant Dirac equation
for $P$, exact to ${\cal O}(\gamma_{\mu\nu})$, is \cite{pap1}
\begin{equation}\label{Psi}
  \Psi(x)=-\frac{1}{2m}\left(-i\gamma^\mu(x){\cal
  D}_\mu-m\right)e^{-i\Phi_T}\Psi_0(x)\equiv \hat{T}\Psi_{0}\,,
\end{equation}
where ${\cal D}_\mu=\nabla_\mu+i\Gamma_\mu (x)$, $\nabla_\mu$ is
the covariant derivative, $\Gamma_{\mu}(x)$ the spin connection
and the matrices $\gamma^{\mu}(x)$ satisfy the relations
$\{\gamma^\mu(x), \gamma^\nu(x)\}=2g^{\mu\nu}(x)$. Both
$\Gamma_\mu(x)$ and $\gamma^\mu(x)$ can be obtained from the usual
constant Dirac matrices by using the tetrad fields $e_{\hat
\alpha}^\mu$ and the relations
\begin{equation}\label{II.2}
\gamma^\mu(x)=e^\mu_{\hat \alpha}(x) \gamma^{\hat
\alpha}=\left(\delta^{\mu}_{\hat{\alpha}}+h^{\mu}_{\hat{\alpha}}(x)\right)\gamma^{\hat{\alpha}}\,,\qquad
\Gamma_\mu(x)=-\frac{1}{4} \sigma^{{\hat \alpha}{\hat \beta}}
e^\nu_{\hat \alpha}e_{\nu\hat{\beta};\, \mu}\,,
\end{equation}
where $\sigma^{{\hat \alpha}{\hat \beta}}=\frac{i}{2}[\gamma^{\hat
\alpha}, \gamma^{\hat \beta}]$. A semicolon and a comma are also
used as alternative ways to indicate covariant and partial
derivatives respectively. We use units $ \hbar = c = 1$, the
signature of $\eta_{\mu\nu}$ is $-2\,,\Phi_T=\Phi_s+\Phi_G$,
\begin{equation}\label{PhiG}
 \Phi_s(x)={\cal P}\int_P^x dz^\lambda
\Gamma_\lambda (z)\,\,\,,
\Phi_{G}=-\frac{1}{4}\int_P^xdz^\lambda\left[\gamma_{\alpha\lambda,
  \beta}(z)-\gamma_{\beta\lambda, \alpha}(z)\right]L^{\alpha\beta}(z)+
  \frac{1}{2}\int_P^x dz^\lambda\gamma_{\alpha\lambda}k^\alpha\,,
\end{equation}
$ L^{\alpha\beta}(z)=(x^\alpha-z^\alpha)k^\beta-
 (x^\beta-z^\beta)k^\alpha$
and $\Psi_0(x)$ satisfies the usual, flat space-time Dirac
equation.
It is convenient to re-write (\ref{Psi}) in the form $ \Psi(x) =
g(x) exp(-ipx)u_{0}(\vec{p})$, where
\begin{equation} \label{g}
g(x)=\frac{1}{2m}\left[\left(\gamma^{\mu}(p_{\mu}+h^{\hat{\alpha}}_{\mu}(x)p_{\hat{\alpha}}+
\Phi_{G,\mu}(x)\right)+m\right]e^{-i\Phi_{T}}\,,
\end{equation}
and
\begin{equation}\label{BPhiGDer}
  \Phi_{G, \mu}=-\frac{1}{2}\int_P^x dz^\lambda (\gamma_{\mu\lambda,
  \beta}-\gamma_{\beta\lambda,
  \mu})p^\beta+\frac{1}{2}\gamma_{\alpha\mu}p^\alpha\,.
\end{equation}
We claim that the transition amplitude for the process of Fig.1 can be calculated by
introducing the generalized four-momentum
\begin{equation}\label{mom}
P_{\mu}= p_{\mu}+\tilde{h}^{\hat{\alpha}}_\mu
p_{\hat{\alpha}}+\tilde{\Phi}_{G,\mu}\equiv
p_{\mu}+\tilde{P}_{\mu}\,,
\end{equation}
for the incoming fermion, as (\ref{g}) itself suggests. The part
that contains the gravitational field is indicated by
$\tilde{P}_\mu$. In (\ref{mom}), $ \tilde{h}^{\hat{\alpha}}_\mu$,
$\tilde{\Phi}_{G,\mu}$ and $\tilde{\Phi}_G$ are quantities that
must be calculated, once the metric is known. They are related to
the Fourier transforms of the corresponding expressions that
appear in (\ref{II.2}), (\ref{PhiG}) and (\ref{BPhiGDer}).
$P_{\mu}$ is not on-shell. In fact
\begin{equation}\label{mass}
P^{\mu}P_{\mu} \equiv m^{2}_e =m^2 +2\left[p^\mu
h_{\mu}^{\hat{\alpha}}p_{\alpha}+ p^\mu \Phi_{G,\mu}\right]\,,
\end{equation}
where $p_{\mu}p^{\mu}=m^2$ because $p_{\mu}$ is the momentum of
the free fermion represented by $ \Psi_{0}(x)$ in (\ref{Psi}). The
transition amplitude is then
\begin{equation}\label{M1}
M_{f\rightarrow f'\gamma} =
-iZe\eta^{\mu\nu}\bar{u}_{0}(\vec{p'})\varepsilon_{\hat{\mu}(\lambda)}\gamma_{\hat{\nu}}g(|\vec{q}|)u_{0}(\vec{p})\,,
\end{equation}
where $\vec{q}\equiv \vec{p}-\vec{p'}-\vec{\ell}$,
$\,\varepsilon^{\hat{\mu}}_{(\lambda)}$ represents the
polarization of the photon, and $Ze$ is the charge of the fermion.
It may be argued that a transition amplitude $ M'_{f\rightarrow f'\gamma} $ must be
added to (\ref{M1}) at $\mathcal{O(\gamma_{\mu\nu})}$ because the contraction
in (\ref{M1}) is in general accomplished by means of $g^{\mu\nu}$ and $g(x)$ contains
a part that is independent of $\gamma_{\mu\nu}$. The transition amplitude $M'$, estimated in
\cite{PRD82}, is given
by \cite{papvall}
\begin{equation}\label{M2}
M'_{f\rightarrow
f'\gamma}=-iZe\gamma^{\mu\nu}(|\vec{q}|)\bar{u}_{0}(\vec{p'})\varepsilon_{\hat{\mu}(\lambda)}
\gamma_{\hat{\nu}}u_{0}(\vec{p})\,.
\end{equation}
However $M'_{f\rightarrow f'\gamma}$ contains the part $p_{\mu}$ of
(\ref{mom}), that comes from $\Psi_{0}$, but not the new part that
contains the gravitational contribution due to the propagation of
the fermion in the field of the source. To $\mathcal{O(\gamma_{\mu\nu}})$ this process is indistinguishable
from bremsstrahlung and is dealt with,
more properly, in that context.
It will not be discussed further in this work.

The calculation now requires that a metric be selected.

Let us consider the particular instance of a fermion that is
propagating with momentum $p^3 \equiv p$, impact parameter $b\geq
R$ and $x_{2}=0$, from $x_3 =-\infty$ toward a gravitational
source of mass $M$ and radius $R$ placed at the origin and
described by the metric $\gamma_{00}=2\phi\,,
\gamma_{ij}=2\phi\delta_{ij} $, where $\phi=-\frac{GM}{r}$. This
metric is frequently used in lensing problems
\cite{lensing},\cite{pap4}. One finds $ \Gamma_{0}=-1/2 \phi,_j
\sigma^{0j}\,,\Gamma_{i}=-1/2 \phi,_{j} \sigma^{ij}$ and $
e^0_{\hat{i}}=0\,,
 e^0_{\hat{0}}=1-\phi\,,
 e^l_{\hat{k}}=\left(1+\phi\right)\delta^l_k $.
All spin matrices are now expressed in terms of ordinary, constant
Dirac matrices. We also assume that the on-shell conditions $
p'_{\mu}p'^{\mu}=m^2\,,\ell_{\mu}\ell^{\mu}=0$ remain valid.
Extension of the calculation to include different particles, or
higher order gravitational contributions to $p'\,,\ell$ and
(\ref{Psi}) can be derived to all orders in $\gamma_{\mu\nu}$ \cite{MPLA}. The
Fourier transforms of the quantities that appear in (\ref{g}) must
now be calculated. We obtain
\begin{equation}\label{h}
h^{\hat{\alpha}}_{0}(q)p_{\alpha}=8\pi^2\delta(q_0)\delta(q_x)\delta(q_y)p_0
GM K_0(bq_z)\,, h^{\hat{\alpha}}_{3}(q)p_{\alpha}
=8\pi^2\delta(q_0)\delta(q_x)\delta(q_y)p GM K_0(bq_z)\,,
\end{equation}
\begin{equation}\label{phgmu}
\Phi_{G,2}(q)=0\,,\Phi_{G,3}(q)=
-8\pi^2\delta(q_0)\delta(q_x)\delta(q_y)\left(\frac{p^2_0}{p}+p\right)GM
K_0(bq_z)\,.
\end{equation}
Four-momentum conservation to zeroth order only is required
because (\ref{h}), (\ref{phgmu}) are already of
${\cal O}(\gamma_{\mu\nu})$. Though the gravitational field
selected above is stationary, energy conservation must be introduced
because the energy contribution of the field is contained in the generalized momentum
of $P$. We further approximate the Bessel
function $K_0(bq_z)\simeq \sqrt{\pi/2bq_z}e^{-bq_z} [1-1/8bq_z
+...]$, itself a distribution, by $K_0(bq_z)/\pi\simeq 2\pi\delta(bq_z)$
and eliminate $\delta^4(q)$ from (\ref{h}) and (\ref{phgmu}). Conservation of energy-momentum will reappear as a
factor $(2\pi)^4 \delta^{4}(q)$ in the expression for the radiated
power $W$ defined below.
By removing $(2\pi)^4\delta^{4}(q)$ from $h^{\hat{\alpha}}_{\mu}$ and
$\Phi_{G,\mu}(q)$ we obtain  $ \tilde{h}^{\hat{\alpha}}_\mu$ and
$\tilde{\Phi}_{G,\mu}$ of (\ref{mom}). We find
\begin{equation}\label{mom1}
P_0 \simeq p_0 +\frac{GM}{b}p_0=p_0+\tilde{P}_0\,, P_1\simeq
\frac{GM}{b}\left(\frac{p_0^2}{p}+p\right)=\tilde{P}_1\,,
P_2=0\,,P^3\equiv P=p- \frac{GM}{b}p= p+\tilde{P}\,.
\end{equation}
We calculate the power radiated as photons in the process of Fig.1
according to the formula \cite{renton}
\begin{equation}\label{W}
W=\frac{1}{8(2\pi)^2}\int
\delta^{4}(P-p'-\ell)\frac{|M|^2}{Pp'_0}d^3p' d^3\ell\,.
\end{equation}
There are two ways to calculate $|M_{f\rightarrow f'\gamma}|^2$.
In the first one we replace $p_{\alpha}$ with $P_{\alpha}$ in the
field-free ($\gamma_{\mu\nu}=0$) expression given by $\Sigma |M|^2
= Z^2 e^2 [-4m^2(p'_{\alpha}p^{\alpha})+8(p_{\alpha}p^{\alpha})]$.
The gravitational contribution to $M$ then appears in
$\tilde{P}_{\mu}$ exclusively. We also remove the terms $-32m^2
(p'_{\alpha}p^{\alpha})+64m^2$ that do not contain gravitational
contributions and therefore refer to the kinematically forbidden
transition. This yields, to ${\cal O}(\gamma_{\mu\nu})$, the
expression
\begin{equation}\label{M1sq}
\Sigma |M_{f\rightarrow f'\gamma}|^2
=Z^2e^2\left[-4(p'_{\alpha}\tilde{P}^{\alpha})+8
(p_{\alpha}\tilde{P}^{\alpha})\right]\,.
\end{equation}
In a second, alternate approach, we calculate $|M|^2$ directly
($\gamma_{\mu\nu}\neq 0$) from (\ref{M1}). By summing over final
spins and averaging over initial spins and polarizations, we
obtain
\begin{equation}\label{TR}
\Sigma |M_{f\rightarrow f'\gamma}|^{2} =
\frac{Z^2e^2}{2(2m)^2}Tr\left\{(\mathbf{p}' + m)\gamma_{\beta}\left[\left(\mathbf{p}+ \tilde{\mathbf{P}} +
m \right)\left((\mathbf{p} + m)^2 +
(\mathbf{p} + m)\mathbf{\tilde{P}^{*}}+ H (\mathbf{p} +  m)\right)\right] \gamma^{\beta}\right\}\,,
\end{equation}
where $ \mathbf{a} \equiv \gamma^{\mu}\, a_{\mu} $ and $H=p^0/p^3 \phi (\gamma^3
p^0 -\gamma^0 p^3)+(p^0/p^3 \gamma^1 p^0 -\gamma^1
p^3)4GMK_1(bq_z)$. On carrying out the traces of the Dirac
matrices, the contribution from $H$ vanishes.
By further eliminating from
$|M|^2$ the terms that refer to the kinematically forbidden
transition, we again find (\ref{M1sq}).
This supports our claim that the generalized
momentum $P_{\mu}$ introduced in (\ref{mom}) leads to the correct
value of the transition probability by the substitution of
$p_{\mu}$ with $P_{\mu}$ in the field-free expression.
The integration over $d^3p'$ in (\ref{W}) is performed by means of
the identity
\begin{equation}\label{id}
\int \frac{d^3p'}{2p'_0}=\int d^4p'\Theta(p'_{0})\delta(p'^2
-m^2)\,,
\end{equation}
while that over $\theta$, the angle between $\vec{P}$ and $\vec{\ell}$, can
be carried out by writing the
on-shell condition for $p'$ in the form
\begin{equation}\label{shell}
\delta(2|\vec{P}||\vec{\ell}|\cos\theta-P^{\alpha}P_{\alpha}
+2P_0\ell_0 + m^2)\,.
\end{equation}
We find
\begin{equation}\label{W1}
W=\frac{Z^2e^2}{4\pi}\left(\frac{GM}{b}\right)\frac{p_{0}^{2}+p^2}{p^2}\ell^2
\,.
\end{equation}
The radiation spectrum is given by
\begin{equation}\label{spec}
\frac{dW}{d\ell}=\frac{Z^2e^2\ell}{2\pi}\left(\frac{GM}{b}\right)\frac{p_{0}^2 +p^2}{p^2}
\,.
\end{equation}
Equation (\ref{shell}) and the condition
\begin{equation}\label{cos}
|\cos\theta|\approx \frac{1}{2p\ell \sqrt{1-\frac{2GM}{b}}}\left\{\frac{2GM}{b}\left(p_{0}^2 + p^2\right)-
2\ell^2 p_{0}\left(1 + \frac{GM}{b}\right)\right\} \leq 1\,,
\end{equation}
lead, for $p>m$, to the inequality
$ p(GM/b)<\ell\leq p$.
It then follows that the hardest photons are emitted in
the backward direction with energy $\ell \sim
p$ and power
\begin{equation}\label{Wm}
W_{p>m}\sim \frac{2Z^2 e^2 p^2}{8\pi}(\frac{GM}{b})\,,
\end{equation}
which would obviously take its highest values in the
neighborhood of a very compact source.

For $p<m$ the inequality (\ref{cos}) is satisfied for $ \ell(Gm/b)\geq m(1-p/m)$ and we also find
\begin{equation}\label{W4}
W_{p<m}\sim \frac{Z^2e^2}{4\pi}\left(\frac{GM}{b}\right)\frac{m^2\ell^2}{p^2} \left(1+\frac{2p^2}{m^2}\right
)\,,
\end{equation}
which diverges for small values of $p$. This infrared
divergence is well-known and arises as a consequence of the finite energy resolution
$\Delta\epsilon$ of the outgoing fermion. The process, as
calculated, is in fact indistinguishable from that in which
massless particles with energy $\leq \Delta\epsilon$ are also emitted and
from processes in which vertex corrections are present (virtual
massless particles emitted and reabsorbed by the external lines of Fig.1).
When these additional diagrams are calculated all infrared
divergences disappear \cite{weinberg}. In the particular case at
hand, $p$ in (\ref{W4}) can be simply replaced by $ GMp_{0}/b$. Below this value
the process is not kinematically allowed.

The process discussed in this section may be considered as the decay of a fermion of effective
mass $P_{\alpha}P^{\alpha}$ into a photon and a fermion of mass $m$ with a lifetime
 $\tau=m/W$ which is indeed small. Quantitatively, for electrons with $p\sim 1 GeV$, $\ell\sim p$,
in the neighborhood of a canonical neutron star, we find $\tau \sim 3\cdot 10^{-19}s$.

\section{The process $\gamma \rightarrow f +\bar{f}$}

Using the replacements $\ell\rightarrow -L$ and $P \rightarrow -q$ in Fig.1,
we can calculate the process by which a
photon produces a
fermion $f$ of momentum $p'$ and an anti-fermion $\bar{f}$ of
momentum $q$ after propagating in a gravitational field.
In addition $Z^2e^2\rightarrow -Z^2e^2$ because of
the presence of the antiparticle. Conservation of energy-momentum
now requires that $L_{\mu}=p'_{\mu}+q_{\mu}$,
while the dispersion
relations are $p'_{\mu}p'^{\mu}=m^2,\,q_{\alpha}q^{\alpha}=m^2$
and the generalized photon momentum is
$L_{\sigma}=\ell_{\sigma}+\Phi_{G,\sigma}$. In the center of mass frame of the
$(f, \bar{f})$-system we now have $\vec{p'}+\vec{q}=\vec{L}=0$ and also $L_{i}=\ell_{i}+\Phi_{G,i}=0$.
This and $\ell_{\alpha}\ell^{\alpha}=0$ again show that if the effect of
the gravitational field vanished ($\Phi_{G}=0$), we would re-obtain the meaningless result
$\ell_{0}=\vec{\ell}=0$. It is therefore the presence of the gravitational field
that enables the process. The effect of the
gravitational field on the polarization of the photon is given by
$E_{\sigma}=\varepsilon_{\sigma}+\frac{i}{2}(\gamma_{\alpha\sigma,\beta}-
\gamma_{\beta\sigma,\alpha})S^{\alpha\beta}
+\frac{i}{2}\gamma_{\alpha\beta,\sigma}T^{\alpha\beta}$ where the
matrices $S^{\alpha\beta}$ and $T^{\alpha\beta}$ are given in
\cite{pap4} and act on the matrix $a_{\alpha}$.
The contributions of these terms reduce to
$\frac{1}{2}(a_{1}^{+}a_{2}-a_{2}^{+}a_{1})(\gamma_{1\sigma,2}-\gamma_{2\sigma,1})$.
In the approximation $K_{0}(b\kappa)/\pi \rightarrow
(2\pi) \delta(b\kappa)$ to be used below, the derivatives of
$\gamma_{\mu\nu}$ behave as $(b\kappa) \delta(b\kappa)=0$ and the
whole contribution of $S^{\alpha\beta}$ and $T^{\alpha\beta}$ to
the photon polarization can be neglected.
We assume that $\ell_{1}=\ell_{2}=0,\,\ell_{3}\equiv \ell$ and
use the following result
\begin{equation}\nonumber
\Phi_{G,1}=\frac{1}{2}\int^{z}_{-\infty} dz^0
\gamma_{00,1}\ell+\frac{1}{2}\int^{z}_{-\infty} dz^3
\gamma_{33,1}\ell=2GM\frac{\ell}{b}(1+\frac{z}{\sqrt{b^2
+z^2}})\,,
\end{equation}
which, for $r\sim b$, becomes $\Phi_{G,1}\sim 2GM\ell/b $.
Similarly, we find
\begin{equation}\nonumber
\Phi_{G,3}=-\frac{GM\ell}{\sqrt{b^2
+z^2}}-GM\ell\int^{z}_{-\infty}dz
\frac{z}{(b^2+z^2)^3/2}=-\frac{2GM\ell}{\sqrt{b^2+z^2}}\sim -\frac{2GM}{b}\ell\,,
\end{equation}
in the same approximation. Factorizing $(2\pi)\delta(b\kappa)$, we find
the generalized momenta $L_0=\ell\,,L_1\sim
2GM\ell/b\,,L_2=0\,,L_3=\ell(1-2GM/b)$. By applying the
substitutions indicated above, we can derive the transition
amplitude for the process $\gamma\rightarrow f+\bar{f}$
\begin{equation}
\Sigma |M^2_{\gamma\rightarrow f\bar{f}}|=Z^2
e^2\left[4(p'_{\alpha}q^{\alpha})+8(q_{\alpha}q^{\alpha})\right]\,,
\end{equation}
and the rate at which energy is radiated as an anti-fermion
\begin{equation}
W_{1}=\frac{1}{8(2\pi)^2}\int \frac{d^3p' d^3q}{q_0p'_0L_0}
\delta^4(q+p'-L)\Sigma |M^2_{\gamma\rightarrow f\bar{f}}|q_0\,.
\end{equation}
The integration over $d^3p' $ can be carried out by means of the
identity (\ref{id}) and the on-shell condition for $p'$ becomes
\begin{equation}\label{on-shell}
\delta\left((L_{\alpha}-q_{\alpha}\right)(L^{\alpha}-q^{\alpha})-m^2)=\delta(L_{\alpha}L^{\alpha}-2L_{0}q^{0}+
2|\vec{L}||\vec{q}|\cos\theta_{1})\,.
\end{equation}
Integrating over $\theta_{1}$, the angle between $\vec{q}$ and $\vec{L}$, we find
\begin{equation}
W_{1}=\frac{Z^2e^2}{4\pi}\int
dq\frac{q}{L_{0}|\vec{L}|}\left(\frac{1}{2}L_{\alpha}L^{\alpha}+m^2\right)\,.
\end{equation}
We also find
$L_{\alpha}L^{\alpha}\simeq2\ell^{\alpha}\Phi_{G,\alpha}\sim-2GM\ell/b$
after replacing $K_{0}(b\kappa)/\pi$ with
$(2\pi)\delta(b\kappa)$, factorizing
$(2\pi)\delta(b\kappa)$ and writing $1/|\vec{L}|\simeq (1+GM/b)/\ell$. We
finally obtain
\begin{equation}\label{faf}
W_{1}=\frac{Z^2e^2}{8\pi}\left(\frac{GM}{b}\right)\frac{m^2q^2}{\ell^2}\left(1-\frac{\ell^2}{m^2}\right)\,,
\end{equation}
from which $\frac{dW_{1}}{dq}$ can be immediately obtained.
We also have $\Theta(L_{0}-q_{0})=\ell-q_{0}>0$, while $|\cos\theta_{1}|\leq 1$ leads to $0\leq \frac{4GM\ell}{b}+ 2q_{0}$ which
is always satisfied in the interval $\pi/2\leq \theta_{1} \leq \pi$.

\section{The process $ f+\bar{f} \rightarrow \gamma $}

We now consider the process in which a fermion-antifermion couple
in the initial state annihilates into a photon. We assume that the
fermion that propagates in the gravitational background has generalized
momentum $P$. By conservation of energy-momentum we then have
$P_{\mu}+q_{\mu}=\ell_{\mu}$ and the generalized momentum is given
by $P_{0}=p_{0}+p_{0}GM/b\equiv
p_{0}+\tilde{P_{0}}\,,P_{1}=P_{2}=0\,,P_{3}=p_{3}(1+GM/b)-p_{0}GM/b\equiv
p_{3}+ \tilde{P_{3}}$. The transition amplitude becomes
\begin{equation}
\Sigma|M|^2_{f\bar{f}\rightarrow\gamma}=Z^2e^2\left[4(q_{\alpha}\tilde{P^{\alpha}})+
8(p_{\alpha}\tilde{P^{\alpha}})\right]\,,
\end{equation}
from which all terms referring to the kinematically forbidden
transition have been eliminated. We also find
\begin{equation}
W_{2}=\frac{Z^2e^2}{(2\pi)^2}\int \frac{d^3q d^3\ell}{q_{0}P_{0}}
\delta^4(P+q-\ell)\left[4(q_{\alpha}\tilde{P^{\alpha}})+8(p_{\alpha}\tilde{P^{\alpha}})\right]\,.
\end{equation}
The integration over $d^3q$ can be easily performed by using the
identity $\int\frac{d^3q}{2q_{0}}=\int d^4q\Theta(q_{0})\delta(q^2-m^2)$.
We find
\begin{equation}
W_{2}=\left(\frac{Ze}{2\pi}\right)^2\int d^3\ell
\Theta(P_{0}-\ell_{0})\frac{1}{P_{0}}\delta\left((P_{\alpha}-\ell_{\alpha})(P^{\alpha}-\ell^{\alpha})-m^2\right)
\left[p_{\alpha}(\ell^{\alpha}-P^{\alpha})+2(p_{\alpha}P^{\alpha})\right]\,.
\end{equation}
The on-shell condition for $q$ becomes
\begin{equation}
\delta(q^2-m^2)=\delta\left((P_{\alpha}-\ell_{\alpha})(P^{\alpha}-\ell^{\alpha})-m^2\right)=
\frac{1}{2\ell|\vec{P}|}\delta\left(\cos\theta_{2}+\frac{P_{\alpha}P^{\alpha}-2\ell_{0}
P_{0}-m^2}{2\ell|\vec{P}|}\right)\,,
\end{equation}
and the integration over $\theta_{2}$ then yields
\begin{equation}\label{W2}
W_{2}= \frac{Z^2e^2}{4\pi}\frac{GM}{b}\int d\ell
\frac{\ell}{P_{0}|\vec{P}|}\left(p_3p_{0}+m^2\right)\,.
\end{equation}
In order to carry out the integration over $\ell$, we first
calculate
\begin{equation}
\frac{1}{P_{0}|\vec{P}|}\simeq\frac{1}{p_{3}p_{0}}\left[1-\frac{1}{2}\left(\frac{2GM}{b}-
\frac{p_{0}}{p_{3}}\frac{GM}{b}\right)\right]\,,
\end{equation}
which must be substituted in (\ref{W2}). The integration over $\ell$ gives
\begin{equation}
W_{2}=\frac{Z^2e^2}{8\pi}\left(\frac{GM}{b}\right)\left[1 -m^2\left(\frac{1}{p_{3}p_{0}}-
\frac{1}{p_{3}^2}\right)\right]\ell^2\,,
\end{equation}
from which we obtain the radiation spectrum
\begin{equation}
\frac{dW_{2}}{d\ell}=
\frac{Z^2e^2}{4\pi}\left(\frac{GM}{b}\right)\left[1 - m^2\left(\frac{1}{p_{3}p_{0}}-
\frac{1}{p_{3}^2}\right)\right]\ell\,.
\end{equation}
The condition $|\cos\theta_{2}|\leq 1$ then requires that
\begin{equation}
\frac{(m^2
+p_{3}p_{0})\frac{GM}{b}}{p_{0}+p_{3}+\frac{p_{0}^2}{p_{3}}\frac{GM}{b}}\leq
\ell \leq \frac{(m^2
+p_{3}p_{0})\frac{GM}{b}}{p_{0}-p_{3}+\frac{p_{0}^2}{p_{3}}\frac{GM}{b}}\,.
\end{equation}
Notice that the process calculated in this section is not the time-reversed
of $\gamma\rightarrow f+\bar{f}$ because in the latter process gravitation is assumed to act on the incoming
photon line and not on the outgoing fermion line.

\section{Conclusions}

External gravitational fields in radiative processes can be included in the calculation of a
transition probability by simply replacing the momentum $p_{\mu}$ of a particle with its generalized version $P_{\mu}$
in the corresponding expression for the zero-field process. The examples given involve spin-$\frac{1}{2}$ and spin-$1$
particles, but the procedure can be extended to any spin. An essential point here is that the dispersion relations
are altered by the external gravitational
field and can be calculated if the corresponding wave equations
can be solved to ${\cal O}(\gamma_{\mu\nu})$, or higher \cite{caipap,pap1,pap2,pap3,pap4}.
It follows, in particular, that kinematically forbidden processes
similar to that of Fig.\ref{fig:Feynman2} become physical and their transition probabilities can be determined.
The calculation of the gravitational contributions are greatly simplified and can be extended to higher order in $\gamma_{\mu\nu}$.
The applications are not confined to fields of
a Newtonian type, but extend to any gravitational field. In this respect, the procedure presented goes beyond the results that apply to external
electromagnetic potentials \cite{jauch}, not only because the metric has in general ten components rather than just four,
but also because time-independence is not a requirement.

The transition amplitudes derived are $\mathcal{O(\gamma_{\mu\nu})}$ to leading order and can therefore be considerably larger
than those normally studied in the literature. The examples given are limited to
a single type of physically relevant metric and we cannot conclude that the resulting
spectra are general and can be used to identify the processes. The actual determination of the spectra
requires the use of metrics specific to the problems studied.
However, the results suggest that particle processes like bremsstrahlung, \v{C}erenkov radiation, or positron production
in the neighborhood
of compact astrophysical objects, or in cosmology, need to be reconsidered.

Space-time has so far been treated as a linear optical medium, though it is by no means clear
what its ultimate properties will be as a result of quantum gravity.
It is not in particular known whether its index of refraction will
remain unaltered in response to high intensity fluxes of particles. There is scope for research on this and
other properties of space-time.

Our final considerations regard
the back-reaction that physical processes may have on the gravitational background.  We show below that the back reaction is not always
negligible and provide an example of how a very small disturbance
in the wave equation can grow rapidly and alter the background gravitational field.

Equation (\ref{Psi}) requires that $\Psi_{0}(x)$ be a solution of the field-free Dirac equation and, of course, of the equation
$(\eta^{\mu\nu}\partial_{\mu}\partial_{\nu}+m^2)\Psi_{0}(x)=0$.
The approximation procedure still holds true, however, when $\Psi_{0}(x)$ satisfies more general equations \cite{caipap0,papb}.
With the addition of a dissipation term, the equation
for $\Psi_{0}$ becomes
\begin{equation}\label{damp}
\left(\eta^{\mu\nu}\partial_{\mu}\partial_{\nu}+m^2 -2m\sigma \partial_{0}\right)\Psi_{0}=0\,,
\end{equation}
where we take $\sigma=\alpha|\langle\Psi_{0}|\hat{T}|\Psi_{0}\rangle|^2 =\alpha(\frac{m}{p_0}\frac{GM}{2b})^2$ \cite{pap1} and $\alpha$
is a dimensionless, arbitrary parameter, $0\leq\alpha \leq 1$, that reflects the coupling strength of the dissipation term.
When we substitute $\Psi_{0}(x)=\exp(m\sigma x_{0})\phi_{0}(x)$ into (\ref{damp}), we obtain
\begin{equation}\label{damp1}
\left[\partial_{0}^{2}-\partial_{z}^{2}+m^{2}(1-\sigma^{2})\right]\phi_{0}(x)=0\,.
\end{equation}
An example of a problem with similar behaviour is offered by a fluid heated from below. For small temperature gradients the fluid conducts the heat,
but as the gradient increases conduction is not sufficient to lead the heat away and the fluid starts to convect.
In realistic problems the exponential growth of $\Psi_{0}$ does not continue indefinitely, but is restricted at
times $x_{0}>\tau \equiv 1/m\sigma$ by nonlinearities or dispersive effects that may have been initially neglected.

The effect of the new solution $\Psi_{0}$ on $W$ can be found as
follows. We first neglect the change $m\rightarrow
m\sqrt{1-\sigma^2}$ in $W$ because in general $\sigma<1$. Setting $m=1$ for simplicity, the
effect of the exponentially increasing term on (\ref{W}) then
amounts to the transformations
$\delta(P_{0}-p_{0}'-\ell_{0})\rightarrow \delta(-2i\sigma
+P_{0}-p_{0}' -\ell_{0})$ and $\frac{1}{p^2}\rightarrow
\frac{1}{\beta^2 (p_{0}-2i\sigma)^2}\simeq
\frac{p_{0}^{2}}{\beta^{2}[(p_{0}^{2}-4\sigma^2)^2
+4p_{0}^{2}\sigma^{2}]}$, where we have used the relation
$\beta=p/p_0$. $W$ has therefore a resonance at $p_{0}=2\sigma$ of
width $4\sigma^2$. Over times $x_{0}>\tau=(\alpha
m)^{-1}(m/p_{0})^{-2}(GM/2b)^{-2} GeV^{-1}$,
$\Psi=\hat{T}\Psi_{0}$ increases exponentially until the
compensating mechanisms mentioned above kick in. For a proton of
energy $p_0\sim 10\,GeV$ in the field of a canonical neutron star
$ \tau\sim 3.5 \times 10^{-21}\alpha^{-1}\,\, s$. Considerably
higher values of $\tau$ can, of course, be obtained for the
lighter fermions. As $\Psi$ grows, so does the energy momentum
tensor associated with it and the gravitational field it
generates, altering, in the process, the gravitational background.

\end{document}